\documentclass[10pt,final,journal]{IEEEtran}
\usepackage[switch]{lineno} 
\usepackage{tabularx} 
\usepackage{booktabs}
\usepackage{multirow}
\usepackage{makecell}

\usepackage{stmaryrd}
\usepackage{graphicx}
\usepackage{amsthm} 
\usepackage{amsmath} 
\usepackage{bm}
\usepackage{graphicx}
\usepackage{epstopdf}
\usepackage{color}
\usepackage{float}
\usepackage[colorlinks=false,linkcolor=black,urlcolor=black,bookmarksopen=true, hidelinks]{hyperref}
\usepackage{bookmark}
\usepackage[flushleft]{threeparttable}
\usepackage{stackengine}
\usepackage{scalerel}
\usepackage{array}

\usepackage{xcolor,soul}

\usepackage{cite}

\ifCLASSINFOpdf
\else
\fi
\usepackage{amsfonts,amssymb}

\usepackage{amsmath}
\usepackage{algorithm}
\usepackage{algpseudocode}
\ifCLASSOPTIONcompsoc
 \usepackage[caption=false,font=normalsize,labelfont=sf,textfont=sf]{subfig}
\else
 \usepackage[caption=false,font=footnotesize]{subfig}
\fi
\begin{document}
%

\title{Semidefinite Programming Two-way TOA Localization for User Devices with Motion and Clock Drift}

\author{Sihao~Zhao, 
		Xiao-Ping~Zhang, \textit{Fellow, IEEE},
        Xiaowei~Cui,
        and~Mingquan~Lu
\thanks{This work was supported in part by the Natural Sciences and Engineering Research Council of Canada (NSERC), Grant No. RGPIN-2020-04661. \textit{(Corresponding author: Xiao-Ping Zhang)}}
\thanks{S. Zhao, X.-P. Zhang are with the Department of Electrical, Computer and Biomedical Engineering, Ryerson University, Toronto, ON M5B 2K3, Canada (e-mail: sihao.zhao@ryerson.ca; xzhang@ryerson.ca).}
\thanks{X. Cui is with the Department of Electronic Engineering,
	Tsinghua University, Beijing 100084, China (e-mail: cxw2005@tsinghua.edu.cn).}
\thanks{M. Lu is with the Department of Electronic Engineering,
	Beijing National Research Center for Information Science and Technology, Tsinghua University, Beijing 100084, China. (e-mail: lumq@tsinghua.edu.cn).}
}

\markboth{}%
{Shell \MakeLowercase{\textit{et al.}}: Bare Demo of IEEEtran.cls for IEEE Journals}
%




\maketitle

\begin{abstract}
In two-way time-of-arrival (TOA) systems, a user device (UD) obtains its position by round-trip communications to a number of anchor nodes (ANs) at known locations. The objective function of the maximum likelihood (ML) method for two-way TOA localization is nonconvex. Thus, the widely-adopted Gauss-Newton iterative method to solve the ML estimator usually suffers from the local minima problem. In this paper, we convert the original estimator into a convex problem by relaxation, and develop a new semidefinite programming (SDP) based localization method for moving UDs, namely SDP-M. Numerical result demonstrates that compared with the iterative method, which often fall into local minima, the SDP-M always converge to the global optimal solution and significantly reduces the localization error by more than 40\%. It also has stable localization accuracy regardless of the UD movement, and outperforms the conventional method for stationary UDs, which has larger error with growing UD velocity.
\end{abstract}



%
\IEEEpeerreviewmaketitle

\section{Introduction}\label{Introduction}
%
%
%
%
\IEEEPARstart{T}{I}ME-of-arrival (TOA), angle-of-arrival (AOA) and received signal strength (RSS) with respect to anchor nodes (ANs) at known coordinates are three most adopted measurements for positioning a user device (UD) in a wireless localization system \cite{shao2014efficient,shi2019blas,luo2019novel,wang2012novel,hu2017robust,an2020distributed}. Localization schemes with imperfect knowledge of the model parameters based on these measurements are also extensively studied \cite{huang2016rss,coluccia2019hybrid,yu2018novel}. Among them, TOA measurement is widely adopted by real-world applications due to its high accuracy \cite{zafari2019survey,lu2019overview,zhao2014kalman,conti2019soft}.

Using round-trip communication, we can have two transmission timestamps and two reception timestamps to obtain two-way TOA measurements. This scheme requires more communications between the UD and the AN, but it is straightforward to implement and will lead to higher localization accuracy due to more TOA measurements. There are plenty of existing methods on two-way TOA localization. Many of them formulate the problem as a maximum likelihood (ML) estimator \cite{bialer2016two,gholami2016tw,zheng2010joint}. The ML estimator has the asymptotic optimality \cite{kay1993fundamentals}, but it is nonlinear and nonconvex for the localization problem. The iterative methods, which linearize the problem by Taylor series expansion, are commonly adopted \cite{foy1976position,borre2007software}. But they require good initialization and may fall into local minima. 
Closed-form methods \cite{chan1994simple,bancroft1985algebraic,zhao2020closed} and multidimensional scaling (MDS)-based approaches \cite{jiang2016multidimensional,wei2009multidimensional,wu2019coordinate} do not require initial guess and can have satisfactory accuracy in small-error conditions. 

In recent years, convex optimization techniques such as semidefinite programming (SDP) have been adopted to solve the localization problem \cite{xu2011source,wang2019convex,zou2020semidefinite,vaghefi2015cooperative,wang2016robust,su2017semidefinite}. They can approximate the ML problem with a convex estimator by relaxation and show desirable performance under large-error conditions \cite{wang2016robust,su2017semidefinite}.

These previous studies on two-way localization all assume that the UD is stationary. This assumption will cause extra position errors if the UD moves. It also hinders the application of their localization methods in moving scenarios such as unmanned aerial vehicle navigation and wearable IoT device localization. The study in \cite{zhao2020optimaltwo} proposes an ML estimator taking the UD movement into account, and presents a Gauss-Newton iterative method to solve the UD position and clock offset. However, it still suffers from the local minima problem when the initial guess is not accurate enough.

In this paper, in order to ensure a globally optimal solution to localize moving UDs with clock drift using the two-way TOA measurements, we develop a new semidefinite programming (SDP) method, namely SDP-M. It relaxes the original nonconvex cost function of the ML method into a convex one. We conduct numerical simulations to evaluate the performance of the proposed SDP-M method in the 3D scene. Results show that the SDP-M always converge to the global minimum, better than the Gauss-Newton iterative method. The localization accuracy of the SDP-M method increases by more than 40\% compared with the iterative method. Compared with the conventional method, which only applies for stationary UDs, the SDP-M has stable localization accuracy regardless of the UD motion.

\section{Problem Formulation} \label{problem}
\subsection{Two-way TOA System}
In a two-way TOA localization system, there are $M$ ANs placed at known positions, and AN \#$i$'s $N$-dimensional coordinate is denoted by $\boldsymbol{q}_i$, $i=1,\cdots,M$. All the ANs are synchronized to a common clock source. One way to achieve synchronization in this system is using multiple timestamp exchanges between ANs \cite{shi2019blas}. The $N$-dimensional position of the UD, denoted by $\boldsymbol{p}$, is the unknown to be determined.

\begin{figure}
	\centering
	\includegraphics[width=0.90\linewidth]{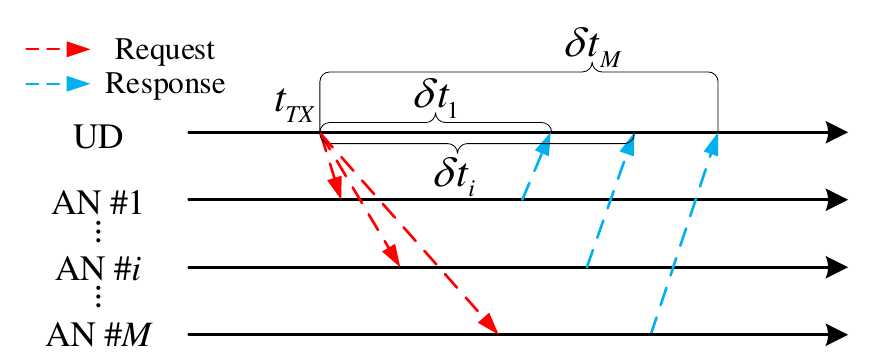}
	\vspace{-0.2cm}
	\caption{Two-way TOA Localization system. The UD transmits the request signal, and all ANs receive. $M$ request-TOA measurements are formed at all ANs. Then, the ANs transmit response signal. The UD receives the response signal with delays. $M$ response-TOA measurements are formed at the UD.
	}
	\label{fig:systemfig}
	\vspace{-0.3cm}
\end{figure}

The two-way TOA measurements are formed through round-trip communications between the UD and the ANs as shown by Fig. \ref{fig:systemfig}. As the UD only transmits one request instead of multiple sequential requests to all the ANs, this scheme can achieve shorter airtime and higher communication capacity. We denote the transmission time of the request signal from the UD by $t_{TX}$, and the interval between the transmission of the request signal and the reception of the response signal from AN \#$i$ by $\delta t_i$, $i=1,\cdots,M$. Without loss of generality, we let the UD first transmit the request signal at $t_{TX}$ and all ANs receive it. By recording the local transmission and reception timestamps, $M$ request-TOA measurements are formed. AN \#$i$ then transmits the response signal received by the UD at $\delta t_i$ to form the response-TOA measurements. After signal transmissions from all the $M$ ANs, we obtain $M$ response-TOA measurements.

We denote the clock offset and drift of the UD with respect to the synchronous ANs by $b$ and $\omega$, respectively. Following the clock model in \cite{hasan2018gnss}, we treat the clock drift as a constant during a short period, and model the clock offset as
\begin{equation} \label{eq:clockbomega}
b(t_2) =b(t_1)+\omega(t_1)\cdot (t_2-t_1)\text{,}
\end{equation}
where $t_1$ and $t_2$ are two time instants close enough to ensure $\omega$ is constant during the interval.

The UD velocity is denoted by $\boldsymbol{v}$. In a short time period, it is reasonable to assume the velocity remains constant. Therefore, we model the UD motion as
\begin{equation} \label{eq:posvel}
\boldsymbol{p}(t_2) = \boldsymbol{p}(t_1) + \boldsymbol{v}(t_1) \cdot (t_2-t_1) \text{.}
\end{equation}

\subsection{TOA Measurement Model}\label{lasproblem}
The request-TOA measurement at AN \#$i$ ($i=1,\cdots,M$), upon reception of the request signal from the UD, is denoted by $\rho_i$. We model it as
\begin{align} \label{eq:rhoANi}
	\rho_i = 
	\frac{\left\Vert\boldsymbol{q}_i-\boldsymbol{p}\left(t_{TX}\right)\right\Vert}{c} -b\left(t_{TX}\right)+ \varepsilon_{i} \text{, } i=1,\cdots,M \text{,}
\end{align}
where $c$ is the signal propagation speed, and $\varepsilon_{i}$ is the measurement noise for AN \#$i$, following independent zero mean Gaussian distribution with a variance of $\sigma_{i}^2$, i.e., $\varepsilon_{i} \sim \mathcal{N}(0,\sigma_{i}^2)$.

Response-TOA measurement, denoted by $\tau_i$, is obtained when the UD receives the response signal from AN \#$i$. Similar to the request-TOA, the response-TOA is related to the true distance between the UD and AN \#$i$ and the clock offset at the instant of reception plus measurement noise. We write the response-TOA $\tau_i$ as
\begin{align} \label{eq:tauANi}
	\tau_i =& \frac{\left\Vert\boldsymbol{q}_i-\boldsymbol{p}\left(t_{TX}\right)-\boldsymbol{v}\left(t_{TX}\right)\cdot\delta t_i\right\Vert}{c}\nonumber\\
	&+ b\left(t_{TX}\right)+\omega\left(t_{TX}\right)\cdot\delta t_i + \varepsilon \text{, } i=1,\cdots,M,
\end{align}
where $\varepsilon$ is the measurement noise for the UD, following a zero-mean Gaussian distribution with a variance of $\sigma^2$, i.e., $\varepsilon \sim \mathcal{N}(0,\sigma^2)$. 

The localization problem for a moving UD is to estimate the coordinate $\boldsymbol{p}$ at the instant $t_{TX}$, given the two-way TOA measurements in (\ref{eq:rhoANi}) and (\ref{eq:tauANi}).

\section{Semidefinite Programming Two-way TOA Localization for Moving UDs} \label{locmethod}

\subsection{ML Estimator for Two-way TOA Localization} \label{estimator}
The unknown parameter we are interested in for localization is the UD position $\boldsymbol{p}$ at the instant $t_{TX}$. However, by observing (\ref{eq:rhoANi}) and (\ref{eq:tauANi}), we also need to handle the UD clock offset $b$, velocity $\boldsymbol{v}$ and clock drift $\omega$. 
The unknown parameter vector is
\begin{align} \label{eq:thetadef}
\boldsymbol{\theta}=
\left[\boldsymbol{p}^T,b,\omega,\boldsymbol{v}^T\right]^T
 \text{.}	
\end{align}


The two-way TOA measurements are written in the collective form as
$$
\boldsymbol{\gamma}=[\boldsymbol{\rho}^T,\boldsymbol{\tau}^T]^T=
\left[\rho_1,\cdots,\rho_M,\tau_1,\cdots,\tau_M\right]^T \text{.} 
$$

The relation between the unknown parameters and the measurements is
\begin{equation} \label{eq:rhoandtheta}
\boldsymbol{\gamma} = h(\boldsymbol{\theta}) + \boldsymbol{\varepsilon} \text{,}
\end{equation}
where based on (\ref{eq:rhoANi}) and (\ref{eq:tauANi}), the $i$-th row of the function $h(\boldsymbol{\theta})$ is
\begin{align} \label{eq:funtheta}
&\left[h(\boldsymbol{\theta})\right]_{i,:} = \nonumber\\
&\left\{
\begin{matrix}
\frac{\left\Vert\boldsymbol{q}_i-\boldsymbol{p}\right\Vert}{c}-b, & i=1,\cdots,M \\
\frac{\left\Vert\boldsymbol{q}_{i-M}-\boldsymbol{p}_{i-M}\right\Vert}{c}+b+\omega \cdot \delta t_{i-M},&i=M+1,\cdots,2M
\end{matrix} 
\right.\text{,}
\end{align}
$\boldsymbol{p}_i=\boldsymbol{p}+\boldsymbol{v}\cdot \delta t_i$, and $\boldsymbol{\varepsilon}=\left[\varepsilon_1,\cdots,\varepsilon_{M},\varepsilon\boldsymbol{1}_M^T\right]^T$ with $\boldsymbol{1}_M$ being an all-one $M$-vector.

Because all the error terms are independently Gaussian distributed, we write the ML estimation of $\boldsymbol{\theta}$ into a weighted least squares (WLS) minimizer as
\begin{equation} \label{eq:MLminimizer}
\hat{\boldsymbol{\theta}}=\text{arg}\min\limits_{{\boldsymbol{\theta}}} \left(\boldsymbol{\gamma} - \mathit{h}({\boldsymbol{\theta}})\right)^T{\bm{W}}\left(\boldsymbol{\gamma} - \mathit{h}({\boldsymbol{\theta}})\right)
\text{,}
\end{equation}
where $\hat{\boldsymbol{\theta}}$ is the estimator, and $\bm{W}$ is a diagonal positive-definite weighting matrix given by
\begin{equation} \label{eq:matW}
\bm{W}=\mathrm{blkdiag}\left(\bm{W}_{\rho},\bm{W}_{\tau}\right)
\text{,}
\end{equation}
in which $\mathrm{blkdiag}(\cdot)$ is a block diagonal matrix, and
\begin{equation} \label{eq:matWrho}
\bm{W}_{\rho}=\mathrm{diag}\left(1/\sigma_1^2,\cdots,1/\sigma_M^2\right),
\bm{W}_{\tau}= \frac{1}{\sigma^2}\bm{I}_M\text{,}
\end{equation}
with $\bm{I}_M$ being an $M\times M$ identity matrix and $\mathrm{diag}(\cdot)$ being a diagonal matrix.


\subsection{Semidefinite Programming for Moving UDs (SDP-M)}
The minimization problem given by (\ref{eq:MLminimizer}) is nonlinear and nonconvex, and it is thus difficult to find a globally optimal solution. In this sub-section, we convert it to a convex problem, namely SDP-M, using semidefinite programming.

The objective function in (\ref{eq:MLminimizer}) is rewritten as
\begin{align} \label{eq:replace1}
\left(\boldsymbol{\gamma} - \mathit{h}({\boldsymbol{\theta}})\right)^T{\bm{W}}\left(\boldsymbol{\gamma} - \mathit{h}({\boldsymbol{\theta}})\right)=\left(\boldsymbol{\gamma} - \bm{A}\boldsymbol{g}\right)^T{\bm{W}}\left(\boldsymbol{\gamma} - \bm{A}\boldsymbol{g}\right)
	\text{,}
\end{align}
where 
$\bm{A}=\left[
\begin{matrix}
	\bm{A}_{\rho}^T &\bm{A}_{\tau} ^T
\end{matrix}
\right]^T
\text{,}
$
in which
\vspace{-0.2cm}
\begin{gather}
\bm{A}_{\rho}=[\bm{I}_M,\bm{O}_{M},-\boldsymbol{1}_M,\boldsymbol{0}_M]\text{, }
\bm{A}_{\tau}=\left[\bm{O}_{M},\bm{I}_M,\boldsymbol{1}_M,\boldsymbol{\lambda}\right], \\
\boldsymbol{\lambda}=[\delta t_1,\cdots,\delta t_M]^T \text{, }
	\boldsymbol{g}=[\boldsymbol{g}_\rho^T,\boldsymbol{g}_\tau^T,b,\omega]^T \text{,}
\end{gather}
with $\bm{O}_{M}$ being a $M\times M$ zero-entry square matrix, and
\begin{gather}
\boldsymbol{g}_\rho=\left[\Vert\boldsymbol{q}_1-\boldsymbol{p}\Vert,\cdots,\Vert\boldsymbol{q}_M-\boldsymbol{p}\Vert\right]^T\text{,}\\
\boldsymbol{g}_\tau=\left[\Vert\boldsymbol{q}_1-\boldsymbol{p}_1\Vert,\cdots,\Vert\boldsymbol{q}_M-\boldsymbol{p}_M\Vert\right]^T \text{.}
\end{gather}

We notice that when we obtain the minimum of the objective function (\ref{eq:MLminimizer}), the partial derivatives with respect to $b$ and $\omega$ equals to zero. This leads us to
\begin{gather}
\frac{\partial\left(\boldsymbol{\gamma} - \bm{A}\boldsymbol{g}\right)^T{\bm{W}}\left(\boldsymbol{\gamma} - \bm{A}\boldsymbol{g}\right)}{\partial b}=0 \Rightarrow \nonumber\\
\left(\bm{A}_{\rho}\boldsymbol{g}-\boldsymbol{\rho}\right)^T\bm{W}_{\rho}\boldsymbol{1}_M+\left(\boldsymbol{\tau}-\bm{A}_{\tau}\boldsymbol{g}\right)^T\bm{W}_{\tau}\boldsymbol{1}_M=0, \label{eq:partialb}\\
\frac{\partial\left(\boldsymbol{\gamma} - \bm{A}\boldsymbol{g}\right)^T{\bm{W}}\left(\boldsymbol{\gamma} - \bm{A}\boldsymbol{g}\right)}{\partial \omega}=0 \Rightarrow\nonumber\\
\left(\boldsymbol{\tau}-\bm{A}_{\tau}\boldsymbol{g}\right)^T\bm{W}_{\tau}\bm{\lambda}=0. \label{eq:partialomega}
\end{gather}

Equations (\ref{eq:partialb}) and (\ref{eq:partialomega}) provide constraints on $b$ and $\omega$, as well as on $\boldsymbol{p}$ and $\boldsymbol{v}$, and improve the localization estimation accuracy.

We note that for a vector $\boldsymbol{x}$, there is $\boldsymbol{x}^T\bm{W}\boldsymbol{x}=\mathrm{tr}\left(\bm{W}\boldsymbol{x}\boldsymbol{x}^T\right)$, where $\mathrm{tr}(\cdot)$ is the trace of a matrix. Therefore, (\ref{eq:replace1}) becomes
\begin{align} \label{eq:replace2}
&\left(\boldsymbol{\gamma} - \mathit{h}({\boldsymbol{\theta}})\right)^T{\bm{W}}\left(\boldsymbol{\gamma} - \mathit{h}({\boldsymbol{\theta}})\right)\nonumber\\
&=\mathrm{tr}\left(\bm{W}\left(\boldsymbol{\gamma}^T\boldsymbol{\gamma} -2\bm{A}\boldsymbol{g}\boldsymbol{\gamma}^T+\bm{A}\bm{G}\bm{A}^T\right)\right),
\end{align}
where $\bm{G}=\boldsymbol{g}\boldsymbol{g}^T$.

We define $y=\boldsymbol{p}^T\boldsymbol{p}$ and $\boldsymbol{z}=[\boldsymbol{p}_{1}^T\boldsymbol{p}_{1},\;\cdots,\;\boldsymbol{p}_{M}^T\boldsymbol{p}_{M}]^T$. The diagonal elements of $\bm{G}$ are 
\begin{align} \label{eq:diagonal1}
[\bm{G}]_{i,i}&=\left(\boldsymbol{q}_i-\boldsymbol{p}\right)^T\left(\boldsymbol{q}_i-\boldsymbol{p}\right)/c^2\nonumber\\
&=\left(\boldsymbol{q}_i^T\boldsymbol{q}_i-2\boldsymbol{q}_i^T\boldsymbol{p}+y\right)/c^2, \; i=1,\cdots,M
\text{,}
\end{align}
\begin{align}\label{eq:diagonal2}
	[\bm{G}]_{i,i}
	&=\left(\boldsymbol{q}_{i-M}^T\boldsymbol{q}_{i-M}-2\boldsymbol{q}_{i-M}^T\boldsymbol{p}_{i-M}+[\boldsymbol{z}]_{i-M}\right)/c^2, \nonumber\\
	i&=M+1,\cdots,2M
	\text{,}
\end{align}

The relation between $\boldsymbol{z}$ and $y$ is
\begin{align}
	[\boldsymbol{z}]_{i}&=y+2\boldsymbol{p}^T\boldsymbol{v}\delta t_i+\boldsymbol{v}^T\boldsymbol{v}\delta t_i^2\nonumber\\
	&=y+\psi\delta t_i+f\delta t_i^2
	\text{, }i=1,\cdots,M,
\end{align}
where $\psi=2\boldsymbol{p}^T\boldsymbol{v}$ and $f=\boldsymbol{v}^T\boldsymbol{v}$.

We use the relations given by (\ref{eq:partialb}), (\ref{eq:partialomega}), (\ref{eq:replace2}), (\ref{eq:diagonal1}), and (\ref{eq:diagonal2}), drop the constant term $\boldsymbol{\gamma}^T\boldsymbol{\gamma}$ in (\ref{eq:replace2}), and the ML problem of (\ref{eq:MLminimizer}) becomes
\begin{gather} \label{eq:reform1}
\min\limits_{\boldsymbol{p},\boldsymbol{v},\boldsymbol{g},\bm{G},\psi,f,y,\boldsymbol{z}} \mathrm{tr}\left(\bm{W}\left( \bm{A}\bm{G}\bm{A}^T-2\bm{A}\boldsymbol{g}\boldsymbol{\gamma}^T\right)\right)
\end{gather}
subject to
\begin{gather}
\left(\bm{A}_{\rho}\boldsymbol{g}-\boldsymbol{\rho}\right)^T\bm{W}_{\rho}\boldsymbol{1}_M+\left(\boldsymbol{\tau}-\bm{A}_{\tau}\boldsymbol{g}\right)^T\bm{W}_{\tau}\boldsymbol{1}_M=0 \label{eq:c1}\\
\left(\boldsymbol{\tau}-\bm{A}_{\tau}\boldsymbol{g}\right)^T\bm{W}_{\tau}\bm{\lambda}=0\label{eq:c2}
\end{gather}
\vspace{-0.9cm}
\begin{align} \label{eq:c3}
&[\bm{G}]_{i,i}=\nonumber\\
&\left\{
	\begin{matrix}
		\boldsymbol{q}_i^T\boldsymbol{q}_i-2\boldsymbol{q}_i^T\boldsymbol{p}+y,& 1\leq i \leq M \\
		\boldsymbol{q}_{i-M}^T\boldsymbol{q}_{i-M}-2\boldsymbol{q}_{i-M}^T\boldsymbol{p}_{i-M}+[\boldsymbol{z}]_{i-M},&M< i\leq2M
	\end{matrix}\right.
\end{align}
\vspace{-0.6cm}
\begin{gather}
[\boldsymbol{z}]_{i}=y+\psi\delta t_i+f\delta t_i^2,\;1\leq i \leq M, \label{eq:c4}\\
[\boldsymbol{g}]_i\geq 0,\;1\leq i \leq M,\label{eq:c5}\\
\bm{G}=\boldsymbol{g}\boldsymbol{g}^T,
y=\boldsymbol{p}^T\boldsymbol{p}\text{, }\label{eq:c7}\\
f=\boldsymbol{v}^T\boldsymbol{v} \text{, }
\psi=2\boldsymbol{p}^T\boldsymbol{v}. \label{eq:c9}
\end{gather}

We can see that the constraints from (\ref{eq:c1}) to (\ref{eq:c4}) are linear with respect to the variables and are thereby convex. However, the constraints in (\ref{eq:c7}) and (\ref{eq:c9}) are still nonconvex.

We apply semidefinite relaxation to convert the nonconvex constraints (\ref{eq:c7}) and (\ref{eq:c9}) to the convex positive semidefinite constraints as 
\begin{gather}
\bm{G}=\boldsymbol{g}\boldsymbol{g}^T\Rightarrow \begin{bmatrix}
	\bm{G} &\boldsymbol{g}\\
	\boldsymbol{g}^T&1
\end{bmatrix} \succeq \bm{O}_{2M+3},\\
y=\boldsymbol{p}^T\boldsymbol{p}\Rightarrow\begin{bmatrix}
	\bm{I}_N &\boldsymbol{p}\\
	\boldsymbol{p}^T&y
\end{bmatrix} \succeq \bm{O}_{N+1},\\
f=\boldsymbol{v}^T\boldsymbol{v}\Rightarrow \begin{bmatrix}
	\bm{I}_N &\boldsymbol{v}\\
	\boldsymbol{v}^T&f
\end{bmatrix} \succeq \bm{O}_{N+1}, \\
\psi=2\boldsymbol{p}^T\boldsymbol{v}\Rightarrow \begin{bmatrix}
	\bm{I}_N &\boldsymbol{p}+\boldsymbol{v}\\
	(\boldsymbol{p}+\boldsymbol{v})^T&y+f+\psi
\end{bmatrix} \succeq \bm{O}_{N+1}.\label{eq:psi}
\end{gather}

The relaxation for the variable $\boldsymbol{\psi}$ in (\ref{eq:psi}) helps to constrain the inner product of the UD position and the velocity, improving the tightness of the SDP method.
	
After all the above relaxations, we convert the nonconvex problem (\ref{eq:MLminimizer}) into a convex estimator as given by
\begin{gather} \label{eq:SPDestimator}
\text{SDP-M: }\min\limits_{\boldsymbol{p},\boldsymbol{v},\boldsymbol{g},\bm{G},\psi,f,y,\boldsymbol{z}} \mathrm{tr}\left(\bm{W}\left( \bm{A}\bm{G}\bm{A}^T-2\bm{A}\boldsymbol{g}\boldsymbol{\gamma}^T\right)\right)
\end{gather}
subject to (\ref{eq:c1}), (\ref{eq:c2}), (\ref{eq:c3}), (\ref{eq:c4}), (\ref{eq:c5}),
\begin{gather}
\begin{bmatrix}
	\bm{G} &\boldsymbol{g}\\
	\boldsymbol{g}^T&1
\end{bmatrix} \succeq \bm{O}_{2M+3} \text{, }
\begin{bmatrix}
	\bm{I}_N &\boldsymbol{p}\\
	\boldsymbol{p}^T&y
\end{bmatrix} \succeq \bm{O}_{N+1},\\
\begin{bmatrix}
	\bm{I}_N &\boldsymbol{v}\\
	\boldsymbol{v}^T&f
\end{bmatrix} \succeq \bm{O}_{N+1} \text{, } 
\begin{bmatrix}
	\bm{I}_N &\boldsymbol{p}+\boldsymbol{v}\\
	(\boldsymbol{p}+\boldsymbol{v})^T&y+f+\psi
\end{bmatrix} \succeq \bm{O}_{N+1}.
\end{gather}

Once we solve the above SDP problem, the estimated position $\boldsymbol{p}$ and velocity $\boldsymbol{v}$ are directly output as the final solution.

\section{Numerical Simulation} \label{simulation}
We create a 3D simulation scene to evaluate the localization performance of the proposed SDP-M method. Eight ANs are placed on the eight vertices of a 600 m$\times$600 m$\times$600 m cubic area. The moving UD is randomly placed in a larger cubic area with the edge length of 700 m. The two cubes share the same center and have parallel surfaces. The 8 ANs are inside the UD cubic area. Hence the cases with UD placed outside the AN region can be tested as well.

One simulation run contains a full period of the round-trip communications between the UD and all the ANs. We set that the UD receives AN \#$i$'s signal at 10$i$ ms after transmission of the request signal. The UD clock offset and drift are set randomly at the start of each simulated period. We set $b\sim \mathcal{U}(0,20)$ $\mu s$, because 20 $\mu s$ is at the level of coarse synchronization error in a regional positioning system. We set $\omega\sim \mathcal{U}(-10,10)$ parts per million (ppm), because this range is at the level of a temperature compensated oscillator (TCXO). The UD velocity is randomly selected, with its norm $\Vert\boldsymbol{v}\Vert$ drawn from $\mathcal{U}(0,60)$ m/s, the yaw angle drawn from $\mathcal{U}(0,2\pi)$, and the elevation angle drawn from $\mathcal{U}(-\pi/2,\pi/2)$. We set the TOA measurement noise $\sigma$ and $\sigma_i$ identical. They both vary from 0.1 m to 10 m with 4 steps. In practice, we can estimate the measurement noise variance by collecting and analyzing the data from the device before operation. At each step, 5,000 times of Monte-Carlo simulations are run. CVX is adopted to solve the SDP problem \cite{cvx,gb08}.

We evaluate the stability of the proposed SDP-M method in comparison with the Gauss-Newton iterative method (iterative method hereinafter) given by \cite{zhao2020optimaltwo}. We use a random position inside the cube with 700 m side length to initialize the iterative method. The  Cram\'er-Rao lower bound (CRLB) of the ML estimator in (\ref{eq:MLminimizer}) following \cite{zhao2020optimaltwo} is used as the localization accuracy benchmark. The three-fold of the theoretical position error derived from CRLB is adopted as the threshold to judge if a failure in localization occurs. This is a loose threshold, but is useful to identify the difference between the proposed SDP-M method and the iterative method.

The success rates of the proposed SDP-M and the iterative method to obtain the correct localization results are listed in Table \ref{table_sucrate}. We can see that the iterative method fails to converge to the correct solution, i.e., the global minimum, for a number of simulation points. On the contrary, for all the simulated points, the new SDP-M method successfully produces the global optimal results, showing its stability and superiority.


%

\begin{figure}
	\centering
	\vspace{-0.1cm}
	\includegraphics[width=0.99\linewidth]{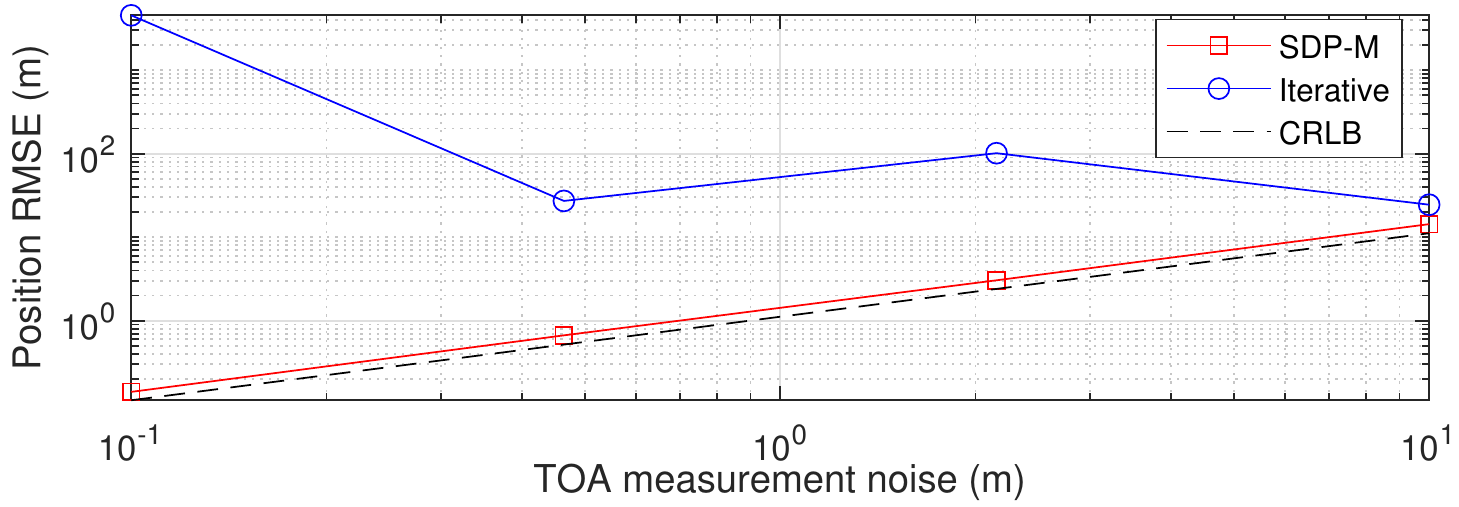}
	\vspace{-0.7cm}
	\caption{Localization error vs. measurement noise for a moving UD. The localization error of the SDP-M is reduced by more than 40\% compared with that of the iterative method.
	}
	\label{fig:presult}
\end{figure}

The localization errors from the SDP-M as well as the CRLB are shown in Fig. \ref{fig:presult}. We can see that the position errors of the SDP-M deviate from the CRLB. This sub-optimality of the new SDP-M method is caused by the relaxation process, which makes the SDP estimator only an approximation of the original ML problem. The localization error of the iterative method is shown in the same figure. Due to the local minima problem, the iterative method produces large error. The localization error of the SDP-M is much smaller, and the improvement compared with the iterative method is more than 40\% for all the noise steps simulated, showing the superiority of the SDP-M.

\begin{table}[!t]
	\centering
	\begin{threeparttable}
		\caption{Success Rate for SDP-M and Iterative Method}
		\label{table_sucrate}
		\centering
		\vspace{-0.3cm}
		\begin{tabular}{c  p{1.8cm}  p{1.8cm} }
			\toprule
			{ Measurement noise $\sigma$ (m)} & {SDP-M}&{Iterative}\\
			
			\hline
			0.10 & 100\%& 83.38\%\\
			0.46 & 100\% &  84.22\%\\
			2.15 & 100\% &  86.20\%\\
			10.00 & 100\% &  96.04\%\\
			\bottomrule
		\end{tabular}
		
		\begin{tablenotes}[para,flushleft]
			Note: The SDP-M localization method gives 100\% globally optimal solution in all simulation runs. The iterative method sometimes fail to obtain the correct solution.
		\end{tablenotes}
	\end{threeparttable}
\vspace{-0.4cm}
\end{table}

%
%

\begin{figure}
	\centering
	\includegraphics[width=0.99\linewidth]{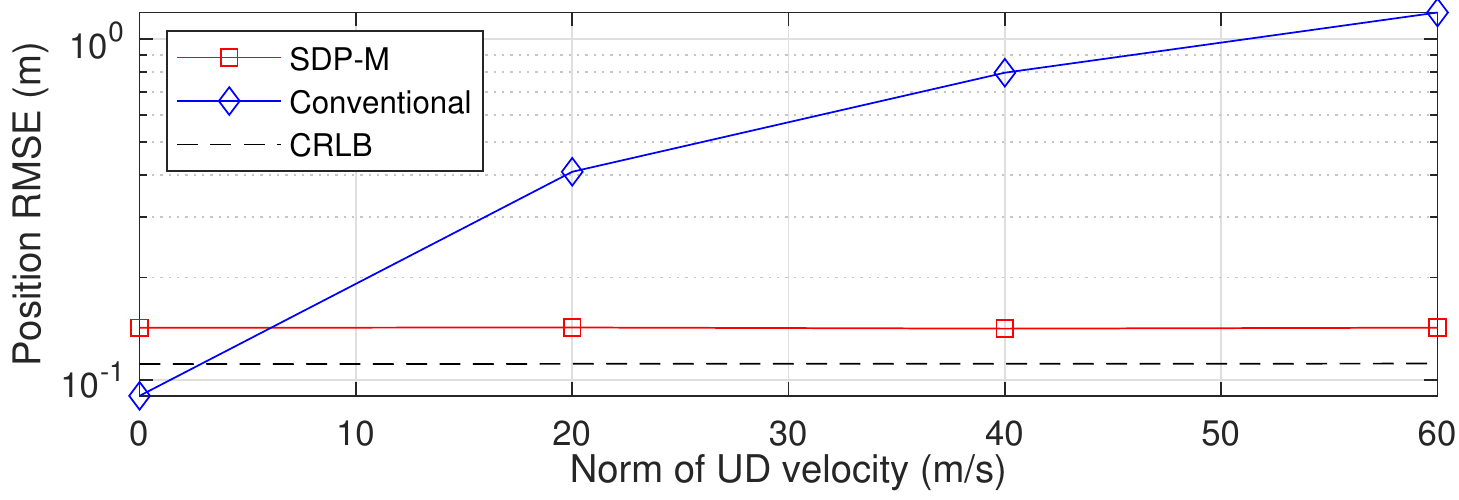} 
	\vspace{-0.7cm}
	\caption{Localization error vs. norm of UD velocity. The localization errors of the SDP-M remain constant with increasing UD speed. The conventional method, which ignores the UD motion, has larger position error with increasing UD speed.
	}
	\label{fig:presultv}
	\vspace{-0.2cm}
\end{figure}

In order to investigate the localization performance of the SDP-M method when the UD velocity changes, we fix the measurement noise to $\sigma$=0.1 m, and vary the speed of the UD from 0 to 60 m/s. The other settings remain the same. The localization error results of the proposed SDP-M and the conventional method, which ignores the UD motion, such as \cite{vaghefi2015cooperative}, are both shown in Fig. \ref{fig:presultv}. The SDP-M method produces stable localization error regardless of the UD speed. The conventional method gives increasing errors, which soon become much larger than that of the SDP-M, when the UD speed grows. The result shows the stable and accurate localization performance of the SDP-M method for a moving UD.

Following the method in \cite{ben2001lectures,wang2016robust}, we estimate the worst case complexity for each iteration of the inner-point algorithm to solve the proposed SDP-M is on the order of $O(M^6)$, where $M$ is the number of ANs, and the iteration count is usually between 20 and 30 \cite{biswas2006semidefinite}. The complexity of the conventional iterative method in one iteration is about $O(M^3)$ \cite{quintana2001note}, and the iteration count limit is set to 10. We record the computation time of 20 simulation runs for each method, and it costs the SDP-M 9.28 s while the iterative method 0.03 s. Since CVX is a universal solver, we expect higher computational efficiency if it can be optimized for the specific problem.

\section{Conclusion}
In this paper, we propose a new semidefinite programming method, namely SDP-M, which relaxes the nonconvex ML-based two-way TOA localization for a moving UD to a convex problem, to ensure the global optimum. Numerical results in a 3D scene with moving UDs show that the new SDP-M method provides the global optimal result. Compared with the iterative method, which may fall into local minima, the new SDP-M always converges to the global minimum, and reduces the localization error by more than 40\%. Compared with the conventional method, which has larger errors with increasing UD velocity, the new SDP-M method gives stable accuracy regardless of the UD motion, showing its superiority.

\ifCLASSOPTIONcaptionsoff
    \newpage
\fi



\bibliographystyle{IEEEtran}
\bibliography{IEEEabrv,paper}
\end{document}